\documentclass[10pt]{article}

\newcommand{\be}{\begin{equation}}
\newcommand{\ee}{\end{equation}}

\begin{document}

\pagenumbering{arabic}

\title{\textbf{\large Sine-Gordon/Coulomb Gas Soliton Correlation Functions and an
Exact Evaluation of the Kosterlitz-Thouless Critical Exponent}}
\author{\normalsize Leonardo Mondaini\footnote{mondaini@if.ufrj.br}\, and E. C.
Marino\footnote{marino@if.ufrj.br}
\\\small{\emph{Instituto de F\'{\i}sica, Universidade Federal do Rio de Janeiro}}}
\date{\small{Cx.Postal 68528, Rio de Janeiro RJ 21941-972, Brazil}}
\maketitle

\begin{abstract}
\hspace{-0.55cm} We present an exact derivation for the asymptotic
large distance behavior of the spin two-point correlation function
in the XY-model. This allows for the exact obtainment of the
critical exponent $\eta=1/4$ at the Kosterlitz-Thouless transition
that occurs in this model and in the 2D neutral Coulomb gas and
which has been previously obtained by scaling arguments. In order
to do that, we use the language of sine-Gordon theory to obtain a
Coulomb Gas description of the XY-model spin correlation function,
which becomes identified with the soliton
correlator of that theory. Using a representation in terms of
bipolar coordinates we obtain an exact expression for the
asymptotic large distance behavior of the relevant correlator at
$\beta^2=8\pi$, which corresponds to the Kosterlitz-Thouless
transition. The result is obtained by approaching this point from
the plasma (high-temperature) phase of the gas. The vortex
correlator of the XY-model is also obtained using the same
procedure.

\end{abstract}

\vskip1.0pc
\section{\normalsize INTRODUCTION}

The sine-Gordon (SG) model is certainly one of the best studied of
$(1+1)$-dimensional physics. The interest in this field
theoretical model has been enhanced by its connections with the
two-dimensional (2D) neutral Coulomb gas (CG) \cite{sgcg} and also
with the 2D XY-magnetic system \cite{KT}. In this framework, it
becomes an useful and powerful tool for the study of a great
variety of physical properties of these two systems, which in
principle admit actual realizations in nature. The SG model is
integrable in the sense that the spectrum and the S-matrix are
exactly known \cite{zz}. Nevertheless, there is a lack of exact
results for correlation functions, except for a specific value of
the coupling parameter \cite{le}.

Many interesting results have been obtained recently, concerning
the SG system. The thermodynamic Bethe ansatz has been used for
obtaining the free energy and specific heat of the system
\cite{dv}. Exact form factors of the soliton operators and other
fields have been derived \cite{sr1,sr2}. Density correlation
functions have been calculated using these form factors
\cite{sr6}. Several results concerning the thermodynamics of the
classical CG have also been obtained \cite{sr7}. Among these, we
mention the exact free energy for $\beta^2 < 4\pi$, in the case of
point particles \cite{sr4} and for $4\pi < \beta^2 < 6\pi$, in the
case of extensive ones \cite{sr5}. Charge and particle correlators
have been obtained in the low temperature ($\beta^2 > 8\pi$) phase
\cite{sr3}.

In this work we present a computation of the two-point spin
correlation function of the XY-model using its connection to the
SG theory. In particular, we exploit the relation existing between
the spin operators of the former and the soliton creation
operators of the latter. Then, we use the CG representation of
these functions in order to derive an exact series describing the
large distance behavior of them. This series clearly exhibits two
distinct types of asymptotic behavior, separated by a critical
point at $\beta^2=8\pi$. At this point, we obtain an exact result
for the spin-spin correlator of the XY-model, which exhibits a
power-law behavior with exponent $\eta=\frac{1}{4}$, well-known
from the scaling analysis of the XY-model \cite{KT}. This special
point, indeed, is the Kosterlitz-Thouless (KT) critical point
\cite{KosT} and, in the 2D CG language, it corresponds to the
temperature $T_c$ in which the system undergoes a phase transition
from a metallic (or plasma) phase, composed of charged particles,
into an insulating (or dieletric) phase, composed of neutral
dipoles (bound pairs of charges). For $\beta^2 < 8\pi$
(high-temperature), our series exhibits a nontrivial large
distance behavior, whereas for $\beta^2 > 8\pi$ (low-temperature),
its large distance behavior is determined by the free theory, as
expected.

The KT critical exponent is usually obtained through scaling
arguments, which are approximate, in the low-temperature phase
\cite{KT,jose,amit,giamarchi} or by heuristic arguments in the
high-temperature phase \cite{minn}. The result presented here is
the first exact evaluation of the KT critical exponent, performed
by approaching the critical point from the high-temperature (in CG
language) phase.

In section 2, we review the connection between the SG theory with
the 2D CG and also with the magnetic XY-model. We pay special
attention to the correspondence between the soliton creation
operators in the SG theory and the spin operators in the 2D
XY-model. This allows us to establish the equivalence between
their respective correlation functions, which can be written in a
convenient form with the help of the 2D CG picture. In section 3,
we make use of bipolar coordinates in order to put the correlation
functions in a form that easily allows for the obtainment of an
exact asymptotic solution at the KT critical point. In section 4,
we consider the case of the vortex correlation function. Finally,
some concluding remarks are presented in section 5. Two appendices
are included. In Appendix A, we review the details of the system
of bipolar coordinates employed in the calculations, whereas in
Appendix B, we demonstrate properties of the functions 
$K\left(|\vec x-\vec y|\right)$ and $K'\left(|\vec x-\vec y|\right)$ 
appearing in our final expressions for the soliton and the
vortex correlation functions, respectively, which are used in
obtaining the asymptotic behavior of these correlators.

\vskip1.0pc
\section{\normalsize SOLITON AND SPIN OPERATORS}

In this section, we are going to review the connection of the SG
theory with the 2D neutral CG and with the 2D XY-model. We will
highlight, in particular, the relation of the soliton creation
operator of that theory with the spin operator of the latter
model, as well as the representation of its correlation functions
in the framework of the classical CG. This will be our starting
point for the evaluation of correlation functions at the KT point.

We start from the SG euclidean action, given by
\begin{equation}
S = \int dx d\tau \left [ \frac{1}{2}\partial_\mu\phi
\partial_\mu\phi +2\alpha_0 \cos\beta\phi \right ]. \label{0}
\end{equation}
It is well-known that the vacuum functional of the theory may be
written as \cite{sgcg,amit}
\begin{equation}
\mathcal{Z}=\lim_{\varepsilon\rightarrow
0}\sum_{n=0}^{\infty}\frac{\alpha^{2n}}{(n!)^2}\int\prod_{i=1}^{2n}d^2z_i\,
\exp\left\{\frac{\beta^2}{8\pi}\sum_{i\neq
j=1}^{2n}\lambda_i\lambda_j\ln \left[|\vec z_i-\vec
z_j|^2+\varepsilon^2\right]\right\}, \label{01}
\end{equation}
where $\lambda_i = 1$ for $1 \leq i \leq n$ and $\lambda_i = - 1$
for $n+1 \leq i \leq 2n$  and $\varepsilon$ is a short-distance
regulator, which is needed in the case of point particles or,
equivalently, of a local field theory. The renormalized coupling
$\alpha$ is related to the one in (\ref{0}) by \be \alpha =
\alpha_0 (\varepsilon^2)^{\frac{\beta^2}{8\pi}} . \label{1} \ee In
expression (\ref{01}) we recognize $\mathcal{Z}$ as the
grand-partition function for the classical 2D CG of point
particles with charges $\pm 1$, where $\alpha$ is the fugacity and
$\frac{\beta^2}{2\pi} = \frac{1}{kT_{CG}}$, where $T_{CG}$ is the
Coulomb gas temperature. In obtaining this relation, we are using
the convention of \cite{conv} for defining the 2D Coulomb
potential. In the case of particles with a finite diameter, $d$,
the integration region for the $2n$ $z_i$-integrals in (\ref{01})
must exclude the regions where $|\vec z_i-\vec z_j|< d$.

The theory possesses soliton excitations, bearing a topological
charge \be Q = \frac{\beta}{2\pi}\int dx\,
\partial_x \phi .
\label{2} \ee At the quantum level, the corresponding states are
created by the soliton operator of Mandelstam \cite{mand} \be
\mu(x,\tau) = \exp \left\{ i\  \frac{2\pi}{\beta}
\int_{-\infty}^{x} dz\, \dot{\phi}  (z,\tau)  \right\}. \label{3}
\ee As we will see below, this operator has an interesting
physical interpretation in terms of the associated spin system,
namely, the XY-model. 

The exact soliton mass spectrum is known, as
well as the masses of soliton bound states (breathers), which
occur for $\beta^2 < 4\pi$ \cite{zz}. Their masses are given by \be M_n = 2M
\sin \left (\frac{\pi \xi}{2}n \right ) \ \ ,\ \
n=1,2,...<\frac{1}{\xi} \label{3a} \ee where $\xi =
\frac{\beta^2}{8\pi - \beta^2}$ and $M$ is the soliton mass.

The SG/CG system is closely related to a 2D spin system,
namely the O(2) ferromagnetic Heisenberg model on a square
lattice, or XY-model \cite{KT}. This is described by the
hamiltonian
\begin{equation}
H_{XY}=-J\sum_{\langle ij\rangle}\vec n_i\cdot\vec n_j, \label{4}
\end{equation}
where \cite{KT,tsv}
\begin{equation}
\vec
n_i=\left(\cos\sqrt{\frac{T}{2J}}\ \theta_i\,,\,\sin\sqrt{\frac{T}{2J}}\ \theta_i
\right). \label{4a}
\end{equation}
The sum in (\ref{4}) runs over nearest neighbors and $T$ is the
temperature (note that the temperature in the XY-model, $T$, is
not the same as in the CG, $T_{CG}$). Observe that $|\vec
n_i|^2=1$ and $\theta_i$ is proportional to the angle the spin
$\vec n_i$ makes with the x-axis.

When we take the continuum limit of (\ref{4}) and consider the
contributions of both spin-waves and vortices to the total energy,
we arrive at the hamiltonian \cite{KT}
\begin{equation}
H_{XY}= \int d^2x \left [ \frac{1}{2} \vec \nabla \theta \cdot
\vec \nabla \theta + \kappa \cos \left (2\pi \sqrt{\frac{2J}{T}}
\ \phi \right ) \right ] , \label{5}
\end{equation}
where we have considered that $\theta_i \rightarrow \theta (\vec
x)$ in the continuum limit and $\phi$ is a scalar field related to
$\theta (\vec x)$ by
\begin{equation}
\theta (x,y) = \int_{-\infty}^x dz\, \partial_y\phi(z,y) .
\label{6}
\end{equation}
From (\ref{6}), it follows that
\begin{equation}
\vec \nabla \theta \cdot \vec \nabla \theta = \vec \nabla \phi
\cdot \vec \nabla \phi \label{7}
\end{equation}
and we see, by comparing (\ref{5}) with (\ref{0}), that the
hamiltonian of the XY-model coincides with the euclidean action of
the SG model, provided we make the identifications: $\beta = 2\pi
\sqrt{\frac{2J}{T}}$ and $\kappa = 2 \alpha_0$. It also follows
that the euclidean vacuum functional of the SG theory, the
grand-partition function of the 2D CG and the partition function
of the XY-model are all identified.

In the XY-model, similarly to the case of the CG of extensive
particles, we have a natural short distance cutoff, namely, the
lattice spacing $\tau$ \cite{KT}. Accordingly, the integration
region for the $2n$ $z_i$-integrals in (\ref{01}) must exclude, in
this case, the regions where $|\vec z_i-\vec z_j|< \tau $.

Within this picture, a very interesting and useful connection can
be established between the XY-spin thermal correlation functions
and the SG quantum soliton correlators. Indeed, in view of
(\ref{6}) we see that the SG soliton operator (\ref{3}) can be
written as \be \mu(x,\tau) = \exp \left\{ i\  \frac{2 \pi}{\beta}
\ \theta (x,\tau)  \right\}. \label{8} \ee The relevant
XY-correlator, therefore, may be written as
\begin{equation}
\langle\vec n(\vec r)\cdot\vec n(\vec 0)\rangle_{XY}=
\langle\mu(\vec r)\mu^\dag(\vec 0)\rangle_{SG}, \label{9}
\end{equation}
where the first expression is the XY-thermal correlator , whereas
the second one is the SG quantum soliton correlator.

Within the CG description, the soliton correlator (\ref{9}) is
given by \cite{amit}:
\begin{eqnarray}
\langle\mu(\vec x)\mu^\dagger(\vec y)\rangle = \frac{\mathcal{Z}^{-1}}{|\vec
x-\vec y|^\frac{2\pi}{\beta^2}}
\sum_{n=0}^{\infty}\frac{\alpha^{2n}}{(n!)^2}\int_{V(r)}\prod_{i=1}^{2n}d^2z_i\,
\nonumber\\  \times \exp \left \{\frac{\beta^2}{8\pi}\sum_{i\neq
j=1}^{2n}\lambda_i\lambda_j\ln |\vec z_i-\vec
z_j|^2+i\sum_{i=1}^{2n} \lambda_i\lbrack \arg(\vec z_i-\vec
y)-\arg(\vec z_i-\vec x)\rbrack \right \}, 
\label{10}
\end{eqnarray}
where $\mathcal{Z}^{-1}$ is given by (\ref{01}).
In the previous expression, we are considering either the case of the
XY-model or the CG of extensive particles, where, as we have
mentioned, there is a natural short distance cutoff. Consequently,
the integrals in (\ref{10}) and in $\mathcal{Z}$ 
are defined in the region $V(r)$, in
which $r < |\vec z_i-\vec z_j|< R $, $r$ being the lattice spacing
$\tau$ in the former case and the particle diameter $d$ in the
latter, whereas $R$ is the radius of the system. $V(r)$, therefore
defines a finite ``volume'' ($V \simeq \pi R^2$ since $r << R$) 
for the system. We have
droped the regulator $\varepsilon$ from  the Coulomb potential
used in (\ref{10}) and in $\mathcal{Z}$, since it is no longer needed in the presence
of the natural short distance cutoff $r$.

Observe that, in (\ref{10}) the contribution coming from the
soliton operators at $\vec x$ and $\vec y$ corresponds, in the CG
language, to the interaction of the charges of the gas with an
external string of electric dipoles orthogonal to it, going from
$\vec x$ to $\vec y$, plus the self-interaction of this string,
which consists in the first term in (\ref{10}) \cite{ms}.

\vskip1.0pc
\section{\normalsize THE SOLITON CORRELATION FUNCTION}

In this section, we are going to obtain a representation of the
soliton correlator (\ref{10}), valid for $\vec x \neq \vec y$,
which will enable us to derive an exact expression for the
asymptotic large distance behavior of
$\langle\mu\mu^\dagger\rangle $ at the KT point, $\beta^2 = 8\pi$.

Indeed, for $\vec x \neq \vec y$ we may use the bipolar
coordinates described in the Appendix A, and rewrite (\ref{10}) as
\begin{eqnarray}
\langle\mu(\vec x)\mu^\dagger(\vec y)\rangle = \frac{\mathcal{Z}^{-1}}{|\vec
x-\vec y|^\frac{2\pi}{\beta^2}}
\sum_{n=0}^{\infty}\frac{\alpha^{2n}}{(n!)^2} \int_{0,V(r)}^{2\pi}
\int_{-\infty,V(r)}^{+\infty} \prod_{i=1}^{2n} d\xi_i  d\eta_i\,
\frac{|\vec x  - \vec y|^{4n}}{4 [\cosh \eta_i -\cos \xi_i]^2
}\nonumber\\ \times \exp \left \{\frac{\beta^2}{8\pi}\sum_{i\neq
j=1}^{2n}\lambda_i\lambda_j \ln \left \{ |\vec x  - \vec y|^2
\left[ \left ( \frac{\sinh \eta_i}{2 [\cosh \eta_i -\cos \xi_i]}-
\frac{ \sinh \eta_j}{2 [\cosh \eta_j -\cos \xi_j]} \right
)^2\nonumber\right .\right .\right .\\ \left . \left .\left . +
\left ( \frac{\sin \xi_i}{2 [\cosh \eta_i -\cos \xi_i]}- \frac{
\sin \xi_j}{2 [\cosh \eta_j -\cos \xi_j]} \right )^2 \right ]
\right \} +i\sum_{i=1}^{2n} \lambda_i \xi_i \right \}. \label{11}
\end{eqnarray}
In this expression, as before, the symbol $V(r)$ expresses the fact that the
integrations must respect the condition that $r < |\vec z_i-\vec z_j|
< R$. In terms of the $\xi_i$,$\eta_i$ integrals, this implies the
following restriction for the expressions between
round brackets in (\ref{11}), which we call, respectively
$\alpha_{ij} $ and $\beta_{ij}$:
\begin{equation}
\left [  \alpha^2_{ij} + \beta^2_{ij}\right ] \in 
\left [\frac{r^2}{|\vec x - \vec y|^{2}},
  \frac{R^2}{|\vec x - \vec y|^{2}} \right ]
\stackrel{|\vec x-\vec y|>> r} {\longrightarrow} 
\left [\frac{r^2}{4R^{2}}, \frac{1}{4}  \right  ]\ .
\label{11a}
\end{equation}

By inspecting (\ref{11}), we immediately conclude that the
$|\vec x  - \vec y|$-factors decouple from the $\xi_i$,$\eta_i$
integrals. Elementary combinatorics, taking into account the
neutrality of the gas, shows that this factor will appear $n(n-1)$
times in the numerator and $n^2$ times in the denominator.
Combining with the $4n$ contribution coming from the scale factors
of the volume elements, we obtain
\begin{equation}
\langle\mu(\vec x)\mu^\dagger(\vec y)\rangle = \frac{\mathcal{Z}^{-1}}{|\vec
x-\vec y|^\frac{2\pi}{\beta^2}} \sum_{n=0}^{\infty}C_n \left(|\vec
x-\vec y|\right)\  |\vec
x-\vec y|^{\left(2 -\frac{\beta^2}{4\pi}\right)2n}. \label{12}
\end{equation}
In this expression the coefficients $C_n$ are given by the summand
in (\ref{11}) after the $|\vec x-\vec y|$-factors have been
removed. In view of the restriction  on the integration region
imposed by (\ref{11a}), we see that these coefficients depend, in
general, on $ |\vec x-\vec y|$. However, in the large distance
regime, where $ |\vec x-\vec y| >> r$, we infer from the right
part of (\ref{11a}) that the integration region appearing in the
expression of $C_n$ is restricted by constant bounds 
and, therefore, do not depend on $ |\vec x-\vec y|$. Hence, we
conclude that the coefficients $C_n$ are constant in this limit.
This fact is confirmed from an independent point of view in
Appendix B.

From Eq. (\ref{12}) we see that for $\beta^2 = 8\pi$ we have 
the following exact expression, valid for the XY-model and the CG of extensive
particles,
\begin{equation}
\langle\mu(\vec x)\mu^\dagger(\vec y)\rangle = \frac{K \left(|\vec
x-\vec y|\right)}{|\vec
x-\vec y|^{\frac{1}{4}}}  \ \ , \label{14b}
\end{equation}
where $K\left(|\vec x-\vec y|\right) = \mathcal{Z}^{-1} \sum_{n=0}^{\infty}
C_n\left(|\vec x-\vec y|\right)$, at $\beta^2 = 8\pi$.

As we shall prove in Appendix B, $K\left(|\vec x-\vec y|\right)$ is a real 
function with an upper bound equal to one, which is saturated in the large distance 
regime, namely, $K \left(|\vec x-\vec y|\right) 
\stackrel{|\vec x-\vec y| >> r} {\longrightarrow} 1 $. This immediately allows us
to write 

\begin{equation}
\langle\mu(\vec x)\mu^\dagger(\vec y)\rangle \stackrel{|\vec x-\vec y| >> r} {\longrightarrow}
\frac{1}{|\vec x-\vec y|^{\frac{1}{4}}} \ \  . 
\label{14b1}
\end{equation}

Eq. (\ref{14b1}) is the well-known result of Kosterlitz and
Thouless for the XY-spin correlation function at the critical
temperature $ T_{KT} = \pi J$ (remember the relation between the
XY-model temperature $T$ and $\beta$, namely, $T = \frac{8\pi^2
J}{\beta^2}$) \cite{KT}. Notice that for determining the critical
exponent of the static and uniform magnetic susceptibility, only
the large distance behavior of the spin correlator is
needed. Expression (\ref{14b1}) provides an exact result for this.

We remark, at this point, that only the leading term in the large distance
behavior of the spin correlator has been obtained when we took 
the $|\vec x - \vec y| >> r$ limit. In order to get the subleading
logarithmic correction obtained previously \cite{KT,amit}, namely 
\begin{equation}
\langle\mu(\vec x)\mu^\dagger(\vec y)\rangle \stackrel{|\vec
x-\vec y| >> r} {\longrightarrow} 
\frac{C\left (\ln |\vec x-\vec y|\right)^{\frac{1}{8}}}{  |\vec x-\vec y|^{\frac{1}{4}}} ,
\label{14c}
\end{equation}
we should consider the next term in the $\frac{r}{|\vec x-\vec y|} $ expansion.
According to the bound and asymptotic value 
obtained for the function $K(x)$ in Appendix B, however, we
conclude that the constant $C$ in (\ref{14c}) must be
$C = (\ln 2 R)^{-\frac{1}{8}}$ such that 
\begin{equation}
K\left(|\vec x-\vec y|\right) 
 \stackrel{|\vec x-\vec y| >> r} {\sim} 
\frac{\left (\ln |\vec x-\vec y|\right)^{\frac{1}{8}}}
{ \left (\ln 2 R\right)^{\frac{1}{8}}} 
 \stackrel{|\vec x-\vec y| >> r} {\longrightarrow} 1 .
\label{14cc}
\end{equation}

Finally, observe that the exponent $1/4$ comes from the free soliton
correlator, which is the prefactor in  (\ref{11}) and (\ref{12}).
This expresses the well-known fact that the $\cos\beta\phi$
interaction becomes irrelevant at $\beta^2 = 8\pi$. Equivalently,
in XY-model language, we would say that the whole contribution to
the asymptotic behavior of the correlators comes from the
spin-wave term. For $\beta^2 > 8\pi$, we still have the large
distance behavior of the correlators determined by the free theory
(spin-wave term). This fact can be inferred directly from
(\ref{12}), even though some logarithmic corrections could be
produced by summing in $n$. For $\beta^2 \geq 8\pi$, a careful
analysis of the ultraviolet divergences is required
\cite{limasantos} in the local case, where any short distance
regulators must be removed at some point. This, however, has no
effect on our result (\ref{12}), where short distance
singularities are absent due to the presence of natural
regulators.

\vskip1.0pc
\section{\normalsize THE VORTEX CORRELATION FUNCTION}

The large distance behavior of the two-point correlation function
of the vortex creation operators in the XY-model can also be
obtained exactly within our approach, at the KT point. Here,
again, the use of the correspondence with the SG theory is quite
useful. Indeed, from (\ref{4a}) it follows that, in SG language,
this operator is given by
\be \sigma(x,\tau) = \exp \left\{ i\  \frac{\beta}{2} \ \phi
(x,\tau)  \right\}.
\label{15} 
\ee 
We can obtain a CG
representation for $<\sigma \sigma^\dagger>$ analogous to
(\ref{11}). The only differences are the prefactor exponent, which
is now $\beta^2/8\pi$ and the last term, in which $\xi_i$ is
replaced by $\eta_i$ and $i$ by $\beta^2/4\pi$. Following the same
procedure as for the soliton correlation function we obtain, at
$\beta^2 = 8\pi$,
\begin{equation}
\langle\sigma(\vec x)\sigma^\dagger(\vec y)\rangle =
\frac{K'\left(|\vec x-\vec y|\right)}{|\vec x-\vec y|} \ \  , 
\label{16}
\end{equation}
where the function $K'\left(|\vec x-\vec y|\right)$ can be obtained from 
$K \left(|\vec x-\vec y|\right)$ through the same
replacements described above. In Appendix B we show that 
$K'\left(|\vec x-\vec y|\right)$ presents the same bound and asymptotic limit
as $K \left(|\vec x-\vec y|\right)$. 

Thus we have  
\begin{equation}
\langle\sigma(\vec x)\sigma^\dagger(\vec y)\rangle
\stackrel{|\vec x-\vec y| >> r} {\longrightarrow}
\frac{1}{|\vec x-\vec y|} \ \  . 
\label{16b}
\end{equation}

\vskip1.0pc
\section{\normalsize CONCLUDING REMARKS}

An interesting extension of this work, which we are now
considering, is the obtainment of the soliton or XY-vortex
four-point correlation functions at the KT point, using the same
methodology. It would also be interesting to compare 
the large distance behavior of our series
(\ref{11}) with the exact solution of the associated free massive
fermion theory at the Luther-Emery point $\beta^2 = 4\pi$
\cite{mand,col}. This would allow us to determine the coefficients
$C_n$ at this point.  Another interesting issue to be explored would 
be the large distance behavior of the soliton and vortex correlators
in the plasma phase ($\beta^2 < 8\pi$), where an exponential decay
should be recovered. Finally, an important case that deserves 
further investigation is the local SG theory or CG of point
particles for $\beta^2 \geq 8\pi$, where a careful treatment of the 
short distance singularities must be performed.

\vfill
\eject

\vskip1.0pc {\normalsize\bf APPENDIX A} \vskip1.0pc

\renewcommand{\theequation}{A.\arabic{equation}}

\setcounter{equation}{0}

Here we give details about the bipolar coordinates, which we use
for computing the soliton correlation function (\ref{10}). Given
the position vector $\vec r$ in the plane and two points (poles)
at $\vec x$ and $\vec y$, we define the coordinates ($\xi,\eta$)
as \cite{arf}
$$
\xi = \arg (\vec r  - \vec y) - \arg (\vec r  - \vec x)
$$
\be \eta = \ln \frac{|\vec r  - \vec x|}{|\vec r  - \vec y|} ,
\label{a1} \ee with $0 \leq \xi \leq 2\pi$ and $-\infty < \eta <
\infty$. In terms of these coordinates, the position vector is
given by \be \vec r =  \frac{|\vec x  - \vec y| }{2 [\cosh \eta
-\cos \xi] }\left (\sinh \eta \ , \
 \sin \xi \right )
\label{a2} \ee and the volume element reads \be d^2z =
\frac{|\vec x  - \vec y|^2}{4 [\cosh \eta -\cos \xi]^2 } d\xi
d\eta. \label{a3} \ee

\vskip1.0pc {\normalsize\bf APPENDIX B} \vskip1.0pc

\renewcommand{\theequation}{B.\arabic{equation}}

\setcounter{equation}{0}

From (\ref{11}), (\ref{12}) and (\ref{14b}) we obtain
\begin{eqnarray}
K \left(|\vec x-\vec y|\right) =  \mathcal{Z}^{-1}
\sum_{n=0}^{\infty}\frac{\alpha^{2n}}{(n!)^2}\int_{V(r)}\prod_{i=1}^{n}d^2x_i
\prod_{i=1}^{n}d^2y_i \frac{\prod_{i<j}^{n} \frac{[x_i ,
x_j]}{|\vec x - \vec y|^4} \prod_{i<j}^{n}  \frac{[ y_i ,
y_j]}{|\vec x - \vec y|^4}}{|\vec x - \vec y|^{4n} \prod_{i,j}^{n}
\frac{[ x_i , y_j]}{|\vec x - \vec y|^4}} \nonumber \\ \times \exp
\left\{i\sum_{i=1}^{n} \lbrack \arg(\vec x_i-\vec y)-\arg(\vec
x_i-\vec x)\rbrack - i\sum_{i=1}^{n} \lbrack \arg(\vec y_i-\vec
y)-\arg(\vec y_i-\vec x)\rbrack\right \}
\label{c1}
\end{eqnarray}
In this expression, we went back to usual coordinates and
associated $\vec x_i$ and $\vec y_i$ with the positive and
negative charges, respectively. We also introduced the symbols 
\be
[ x_i , y_j] \equiv |\vec x_i - \vec y_j|^4 > r^4 .
\label{c2} 
\ee

Let us first prove that $K$ is real. This can be easily done by
taking the complex conjugate of (\ref{c1}) and subsequently performing
the change of variables $x_i \leftrightarrow y_i$.

Considering now the contribution of the phase factors to the integrals in
(\ref{c1}) we can establish the following bound:
\begin{eqnarray}
K \left(|\vec x-\vec y|\right) \leq  \mathcal{Z}^{-1}
 \sum_{n=0}^{\infty}\frac{\alpha^{2n}}{(n!)^2}\int_{V(r)}\prod_{i=1}^{n}d^2x_i
\prod_{i=1}^{n}d^2y_i \frac{\prod_{i<j}^{n} [x_i ,
x_j]\ \  \prod_{i<j}^{n}  [ y_i ,
y_j]}{ \prod_{i,j}^{n}
[ x_i , y_j]} . 
\label{c21}
\end{eqnarray}
Observe that the $|\vec x - \vec y|$-terms in (\ref{c1}) 
have completely canceled out. Notice also that the expression
given by the sum in (\ref{c21}) is nothing but $ \mathcal{Z}$
and therefore we immediately infer that $K \left(|\vec x-\vec y|\right) \leq 1$,
which is the announced bound for the function in (\ref{14b}).

We now consider the asymptotic behavior of $K$, for $|\vec x-\vec y| >> r$.
The phases in (\ref{c1}) cancel out in the leading order in this limit and we
immediately see that the above bound is saturated in the 
leading asymptotic regime, namely \\
$K \left(|\vec x-\vec y|\right) 
\stackrel{|\vec x-\vec y| >> r} {\longrightarrow} 1 $.

In order to complete our analysis, let us establish now an
upper bound for $\mathcal{Z}$ itself. Using the expression
for $\mathcal{Z}$ given by the sum in (\ref{c21}),
we can infer the following bound for
$\mathcal{Z}$, by making the replacement $[ x_i ,  y_j]
\leftrightarrow  r^4 $ in the $n$ $i=j$ terms in the denominator:
\begin{eqnarray}
\mathcal{Z} \leq \sum_{n=0}^{\infty}\frac{\left
(\frac{\alpha}{r^2}\right )^{2n}}{(n!)^2}
\int_{V(r)}\prod_{i=1}^{n}d^2x_i \prod_{i=1}^{n}d^2y_i \
f_n(x_1,...,x_n;y_1,...,y_n) , \label{c3}
\end{eqnarray}
where
\begin{eqnarray}
{f}_n(x_1,...,x_n;y_1,...,y_n) = \frac{\prod_{i<j}^{n} [x_i , x_j]
\prod_{i<j}^{n}  [ y_i ,  y_j]}{\prod_{i\neq j}^{n} [ x_i , y_j]}\
. \label{c4}
\end{eqnarray}
Using the Mean-Value Theorem, we may write the $n$-th integral in
(\ref{c3}) as
\begin{eqnarray}
I_n = V^{2n} \overline{f_n} , \label{c5}
\end{eqnarray}
where $\overline{f_n} $ is the average value of $f_n$ and $V$ is
the ``volume'' ($\simeq \pi R^2$) of the system. Observe that $f_1 = 1$,
whereas $f_2$ is given by
\begin{eqnarray}
f_2 = \frac{[x_1 , x_2][y_1 , y_2]}{[x_1 , y_2][x_2 , y_1]}
\label{c6}
\end{eqnarray}
and
\begin{eqnarray}
f_3 = f_2(x_1,x_2;y_1,y_2) \ f_2(x_1,x_3;y_1,y_3) \
f_2(x_2,x_3;y_2,y_3) . \label{c7}
\end{eqnarray}
The generalization for higher values of $n$ is straightforward
and we have in general $f_n = (f_2)^{\frac{n^2-n}{2}}$.

Consider now $I_2$. Performing the changes of variables $x_1
\rightarrow y_1$ and $y_1 \rightarrow x_1$, we readily conclude
that
\begin{eqnarray}
\overline{f_2} = \overline{f_2^{-1}} . \label{c8}
\end{eqnarray}
This implies that $\overline{f_2} \leq 1$ and also that
$\overline{(f_2)^N} \leq 1$. 

Since, as we have just seen, $f_n $ is in the form $(f_2)^N$
we conclude that $\overline{f_n} \leq 1$
and, consequently, $I_n \leq V^{2n}$. Inserting this bound in
(\ref{c3}), we obtain (notice that $\mathcal{Z}$ is positive)
\begin{eqnarray}
0 < \mathcal{Z}  \leq \sum_{n=0}^{\infty}\frac{\left
(\frac{\alpha \pi R^2}{r^2}\right )^{2n}}{(n!)^2} = {\rm I}_0
\left ( \frac{2\pi \alpha R^2}{r^2} \right )  , 
\label{c9}
\end{eqnarray}
where ${\rm I}_0 (x) $ is a modified Bessel function of the first
kind and we have set $V=\pi R^2$, since $R >> r$. 
From this we see that $\mathcal{Z}$ is finite for the XY-model and the
CG of extensive particles (where there is a natural short distance
cutoff $r$) whenever the volume is finite .

Now let us consider $K'\left(|\vec x-\vec y|\right)$. 
From the explanations given after
(\ref{15}), we conclude that
\begin{eqnarray}
K'\left(|\vec x-\vec y|\right) =  \mathcal{Z}^{-1}
\sum_{n=0}^{\infty}\frac{\alpha^{2n}}{(n!)^2}\int_{V(r)}\prod_{i=1}^{n}d^2x_i
\prod_{i=1}^{n}d^2y_i \frac{\prod_{i<j}^{n} \frac{[x_i ,
x_j]}{|\vec x - \vec y|^4} \prod_{i<j}^{n}  \frac{[ y_i ,
y_j]}{|\vec x - \vec y|^4}}{|\vec x - \vec y|^{4n} \prod_{i,j}^{n}
\frac{[ x_i , y_j]}{|\vec x - \vec y|^4}} \nonumber \\
\frac{\prod_{i}^{n} [x_i ,x]^{1/2} \prod_{i}^{n}  [ y_i ,y]^{1/2}}
{\prod_{i}^{n}  [ x_i ,y]^{1/2} \prod_{i}^{n} [y_i ,x]^{1/2} } .
\label{c10}
\end{eqnarray}

Observe that the last factor in the previous expression 
tends to one in the limit $|\vec x - \vec y| >> r$.
Hence, we immediately conclude,
after the $|\vec x - \vec y|$-terms are canceled out,
that $K' \left(|\vec x-\vec y|\right) 
\stackrel{|\vec x-\vec y| >> r} {\longrightarrow} 1 $,
thus establishing (\ref{16b}).

\vskip1.0pc {\normalsize\bf ACKNOWLEDGMENTS } \vskip1.0pc

This work has been supported in part by CNPq, FAPERJ and
PRONEX-66.2002/1998-9. LM was supported by CNPq and ECM was
partially supported by CNPq.

\end{document}